\documentclass[11pt,aps,preprint,tightenlines,showpacs,groupadress]{revtex4}

\usepackage[english]{babel}
\usepackage[dvips]{graphicx}
\usepackage{indentfirst}
\usepackage{tabularx}
\usepackage[normalem]{ulem}

\begin{document}
\title{Onset temperature of Bose-Einstein condensation in incommensurate solid ${\bf ^4}$He}
\author{R. Rota}
\affiliation{Departament de F\'\i sica i Enginyeria Nuclear, Campus Nord B4-B5, 
Universitat Polit\`ecnica de Catalunya, 08034 Barcelona, Spain}
\author{J. Boronat}
\affiliation{Departament de F\'\i sica i Enginyeria Nuclear, Campus Nord B4-B5, 
Universitat Polit\`ecnica de Catalunya, 08034 Barcelona, Spain}
\date{\today}

\begin{abstract}
The temperature dependence of the one-body density matrix in
$^4$He crystals presenting vacancies is computed with Path Integral
Monte Carlo. The main purpose of this study is to estimate
the onset temperature $T_0$ of Bose-Einstein condensation in these
systems. We see that $T_0$ depends on the vacancy concentration
$X_v$ of the simulated system, but not following the law $T_0 \sim
X_v^{2/3}$ obtained assuming non-interacting
vacancies. For the lowest $X_v$ we have studied, that is $X_v = 1/256$,
we get $T_0 = 0.15 \pm 0.05$ K, close to the temperatures
at which a finite fraction of non-classical rotational inertia is
experimentally observed. Below $T_0$, vacancies do not act as
classical point defects becoming completely delocalized entities.
\end{abstract}


\maketitle

The debate about the supersolid state of matter, i.e., a phase
where crystalline order coexists with superfluidity, has gained a great interest among the scientific community
after the first observation of non classical rotational inertia
(NCRI) in torsional oscillators containing solid
helium~\cite{KimChan04Vycor,KimChan04Bulk}. Although several
experiments have confirmed the appearance of a phase transition in
solid $^4$He at temperatures $T_c \sim 60-100 \, {\rm
mK}$~\cite{Aoki07,Kondo07}, we are still far from a complete description
of this phenomenon because of controversial experimental
results. For instance, the values of the superfluid density
$\rho_s/\rho$ reported so far can vary more than one order of
magnitude according to experimental conditions such as the way in
which the crystal is prepared, its subsequent annealing or the $^3$He
concentration~\cite{Rittner07,Penzev07,Clark07}. These
discrepancies suggest that the quality of the solid sample plays a
very important role in these experiments and make fundamental a
study of crystalline defects in quantum crystals.

First theoretical studies, indeed, suggested that a possible
supersolid behavior can be explained assuming the presence, in the
ground state of quantum crystals, of delocalized vacancies which
may undergo Bose-Einstein condensation (BEC)  at low 
temperature~\cite{AndreevLifshitz69}.
Nevertheless, these early works were based on simplified models,
so that it was not possible to draw specific predictions for solid
$^4$He. More recently, microscopic methods have been extensively used to
provide a reliable description of the supersolid state but, so
far, they have not been able to reproduce all the experimental
findings. Path integral Monte Carlo (PIMC) simulations have shown
that a commensurate perfect crystal does not exhibit 
superfluidity~\cite{Ceperley04,Ceperley06,Galli08}, but a non-zero condensate
fraction has been observed in crystals with a finite vacancy concentration 
at zero temperature~\cite{Galli06}.

The possibility for solid $^4$He to present vacancies in
its ground state seems to be hindered by the energetic cost
of these defects. According to several Quantum Monte Carlo results, the vacancy formation energy is estimated to be of
the order of 10 K~\cite{Pederiva97,Boninsegni06,Clark08,Cazorla09,Yaros10}, in agreement
with experimental measurements~\cite{Fraass89}. Nevertheless, the
high delocalization of the vacancies in solid $^4$He at
temperatures close to zero prevents an interpretation of these defects in terms of a classical
theory involving an activation energy and a configurational
entropy for their creation~\cite{Burns,Anderson05,Anderson09}.
Furthermore, experimental data cannot rule out the possibility of
a zero-point vacancy concentration below 0.4\%~\cite{Simmons07}.
It has also to be noticed that formation energy considerations do
not exclude the possibility of vacancies introduced through the
experimental conditions, for example during the crystal growth.
The spatial correlation between vacancies has
been calculated in order to understand if a gas of defects can be metastable in solid $^4$He. The results show an attractive
correlation between vacancies at short distance, but they cannot
conclude if they form bound states and aggregate in large
clusters which  eventually would phase 
separate~\cite{Boninsegni06,Rossi08,Pessoa09,Lutsyshyn10}.

In this work, we calculate by means of the Path Integral Monte
Carlo (PIMC) method, the one-body density matrix $\rho_1({\bf
r},{\bf r'})$ in solid $^4$He samples presenting a finite
vacancy concentration, focusing especially on its temperature
dependence. In the study of the BEC properties of quantum systems,
$\rho_1({\bf r},{\bf r'})$ is a fundamental quantity, the condensate fraction
$n_0$ being its asymptotic limit for large $|{\bf r} - {\bf r'}|$ values. Our
main purpose is to estimate the onset temperature of BEC $T_0$ and
to compare it with the experimental measurements.
We start simulating an hcp crystal with a vacancy concentration
$X_v = 1/180$, trying also to give a qualitative picture of the
delocalization of the vacancies and the appearance of BEC.
Finally, we study the dependence of $T_0$ on $X_v$ to
guess which would be the vacancy concentration needed to have
BEC appearing in the range of temperatures $T_c \sim 60-100 \,
{\rm mK}$ at which NCRI is experimentally observed.

PIMC provides a fundamental approach in the study of
the thermodynamic properties of strongly interacting quantum
systems at finite temperature~\cite{CeperleyRev}. In this method, the partition 
function $Z$ is rewritten 
making use of the convolution property of the thermal density matrix 
$G(R',R;\beta)=\langle R' \vert e^{-\beta\hat{H}}\vert R \rangle$
(where $\beta = (k_B T)^{-1}$ is the inverse of the temperature
and $\hat{H}$ is the Hamiltonian of the system), which is known only for
small $\beta$. This procedure is equivalent to mapping the quantum many-body system at
finite temperature onto a classical system of closed ring
polymers. Increasing the number of convolution terms used to rewrite $Z$, which 
corresponds to the number of beads composing each classic
polymer, one is able to 
reduce the systematic error due to the approximation for $G$ and therefore to 
recover ``exactly" the thermal equilibrium properties of the system.
A good approximation for the propagator $G$ is fundamental in
order to reduce the complexity of the calculation and ergodicity issues.
Using the Chin approximation~\cite{chin,Sakkos09}, we are able to obtain an accurate 
estimation of the relevant physical quantities with reasonable numeric effort even in the
low temperature regime, where the simulation becomes harder 
due to the large zero-point motion of particles. 
Chin approximation for the action is accurate to fourth order in the
imaginary-time step but a real sixth-order behavior can be achieved by
adjusting properly the two parameters entering in it. Similar accuracies
can be achieved using other high-order proposals for the
action~\cite{predescu,balaz,zillich}.

An additional problem we have to deal with when approaching the
low temperature limit with PIMC simulations arises from the
indistinguishable nature of $^4$He atoms. Since we study a bosonic system, the symmetry of $Z$
can be recovered via the direct sampling of permutations between
the ring polymers. To this purpose, we have used the Worm 
Algorithm~\cite{BoninsegniWorm}. This algorithm allows PIMC simulations with a 
very efficient sampling of the exchanges between bosons. Furthermore, it is able to give 
an estimation of the normalization factor of $\rho_1({\bf r}_1,{\bf r}_1')$, avoiding thus the 
systematical uncertainties which can be
introduced by a posteriori normalization factor.

\begin{figure}
\includegraphics[width=0.7\linewidth]{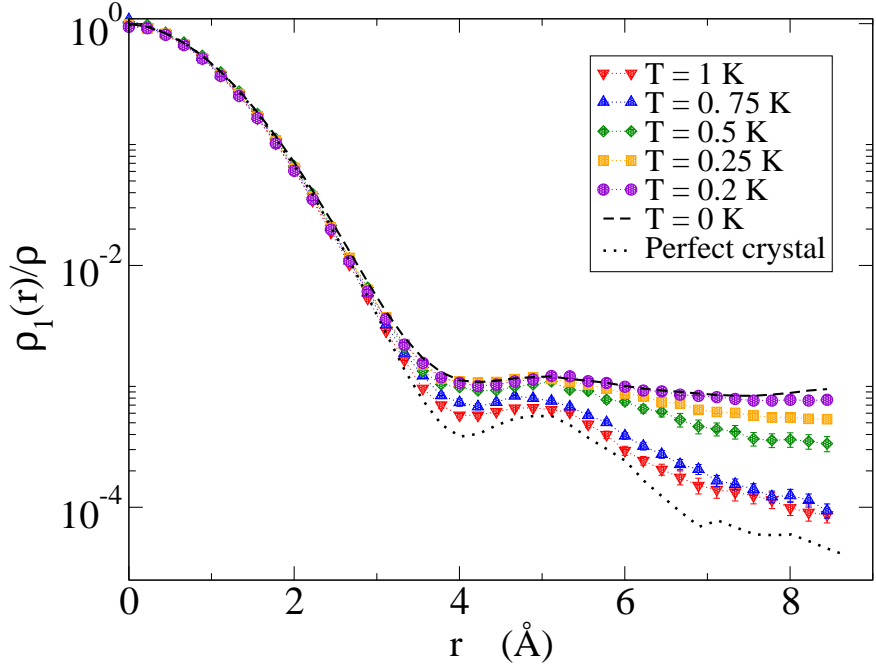}
\caption{(Color online) The one-body density matrix $\rho_1(r)$
for an hcp crystal with vacancy concentration $X_v = 1/180$ at
density $\rho = 0.0294$ \AA$^{-3}$ at different temperatures: $T =
1 \, {\rm K}$ (red triangles down), $T = 0.75 \, {\rm K}$ (blue
triangles up), $T = 0.5 \, {\rm K}$ (green diamonds), $T = 0.25 \,
{\rm K}$ (yellow squares) and $T = 0.2 \, {\rm K}$ (purple
circles). The dotted and dashed lines represent $\rho_1(r)$
at zero temperature respectively for the commensurate ($X_v = 0$)
and incommensurate crystal ($X_v = 1/180$) at the same density,
taken from Ref. \cite{Rota11}.}\label{obdm_vac}
\end{figure}

In order to calculate $\rho_1(r) = \rho_1(|{\bf r} - {\bf r'}|)$ in
a crystal with vacancy concentration $X_v = 1/180$, we have carried out
simulations of $N = 179$ $^4$He atoms, interacting through an accurate
Aziz pair potential~\cite{Aziz}, in an almost cubic simulation box
matching the periodicity of an hcp lattice made up of $N_s = 180$
sites at a density $\rho = 0.0294$ \AA$^{-3}$. We apply periodic
boundary conditions to the simulation box in order to simulate the
infinite dimensions of the bulk system. In Fig. \ref{obdm_vac},
we show the results for $\rho_1(r)$ at different temperatures and
we compare them with the zero temperature estimations of $\rho_1$
for the same system and for a perfect hcp crystal, obtained with
the Path Integral Ground State method in Ref. \cite{Rota11}. We notice
that, at temperatures $T \ge 0.75$ K, $\rho_1(r)$ computed
in an incommensurate crystal, even though is not compatible with
$\rho_1$ for the perfect crystal, presents a similar
exponential decay at large $r$. At lower temperatures, the
decay of $\rho_1(r)$ is smoother and, for temperatures $T \le
T_0 = 0.2$ K, $\rho_1$ presents a non-zero asymptote at large $r$,
which indicates the presence of BEC inside the system. This $T_0$
can be considered a first estimate of the onset temperature of
supersolidity in the simulated system. An analysis of
the finite size effects would be needed to get a more precise
estimation of the critical temperature of the supersolidity
transition. Nevertheless, the simulation of bigger systems with
exactly the same vacancy concentration requires a huge
computational effort that would make the calculations
impracticable.

\begin{figure}
\includegraphics[width=0.625\linewidth]{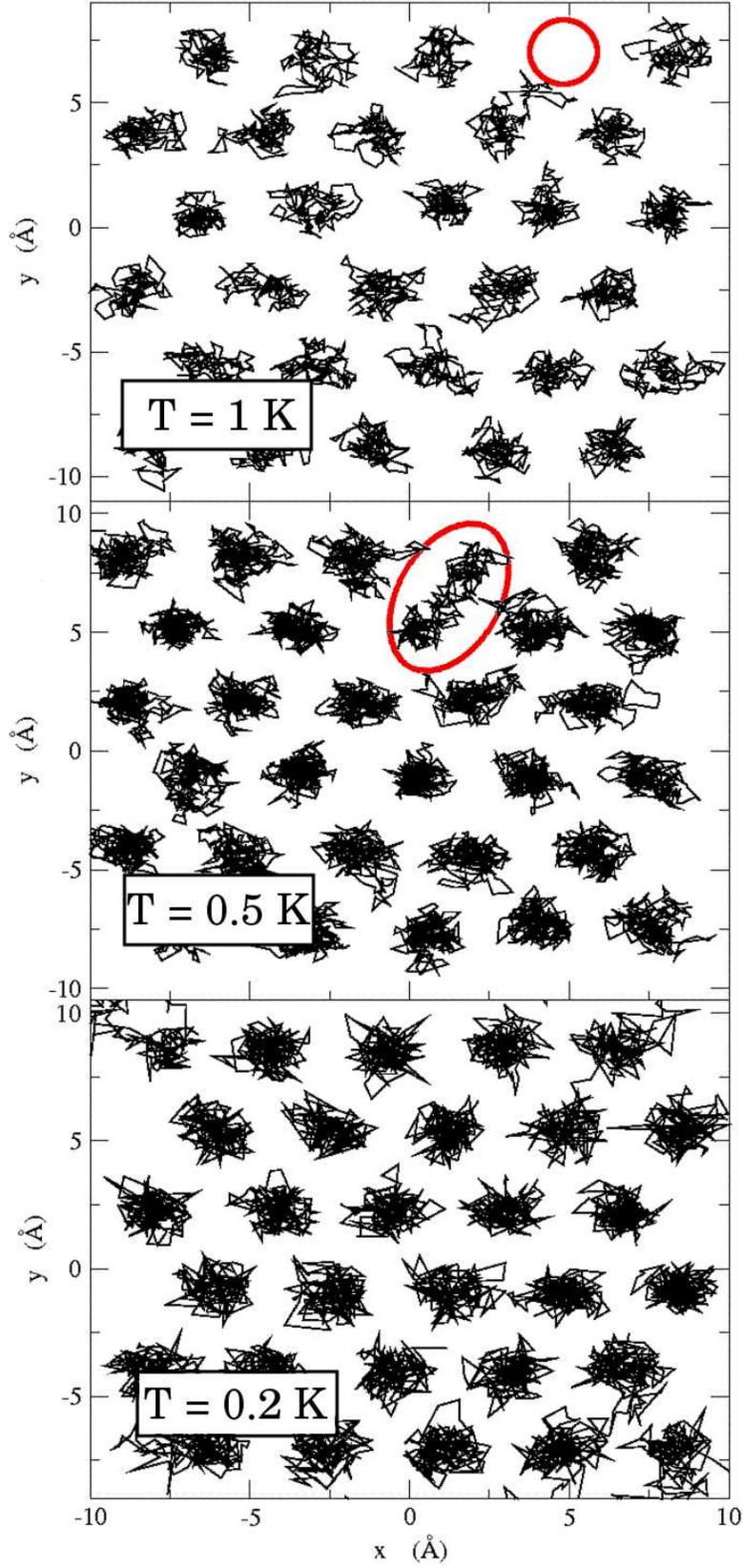}
\caption{(Color online) Two-dimensional projection of basal planes
of the incommensurate hcp crystal at different temperatures,
represented according to the PIMC isomorphism of the classical
polymers. At $T = 1$ K (higher panel) the vacancy is localized and
indicated by the red circle. At $T = 0.5$ K (middle panel), the
vacancy begins to delocalize: the red ellipse indicate a quantum
particle delocalized over two different lattice sites. At $T =
0.2$ K (lower panel), the vacancy is completely delocalized and
cannot be easily detected.}\label{configurations}
\end{figure}

\begin{figure}
\includegraphics[width=0.625\linewidth]{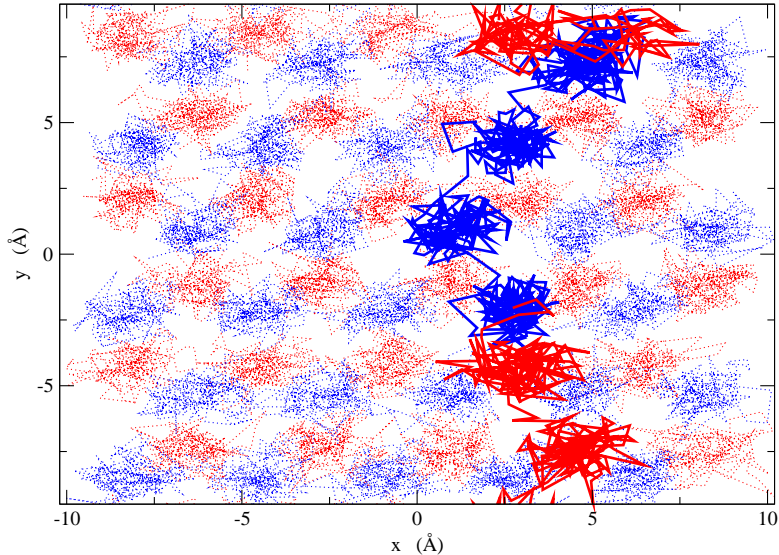}
\caption{(Color online) Two-dimensional projection of two
consecutive basal planes of the incommensurate hcp crystal at $T = 0.15$ K. The different colors distinguish the two different planes. 
The thick solid line represent a long permutation cycle
between the $^4$He atoms presenting a non-zero winding number.}\label{SupersolidPicture}
\end{figure}

In order to give a more qualitative description of the appearance
of BEC in incommensurate $^4$He solids, we visualize typical
configurations of the system during the simulation. In Fig.
\ref{configurations}, we plot two-dimensional
projections of the positions of the quantum particles (represented by
polymers in PIMC) lying in a basal plane of the incommensurate hcp crystal
at different temperature. At $T = 1$ K, $^4$He atoms tend to
stay localized around their equilibrium positions. Also the
vacancies are localized and can be easily detected inside the
lattice. This explains the fact that, at that temperature, the
presence of vacancies does not affect noticeably the overall
behavior of $\rho_1$ which, for the incommensurate crystal, is
similar to the one of the perfect crystal. At $T = 0.5$
K, the effects of the delocalization of the $^4$He atoms can be
seen with the appearance of some polymers which are spread on two
different lattice points. In the space configurations at this
temperature, the acceptance rate of the exchange between the
polymers is higher than in the configurations at larger
temperature, but it is still too low to allow the appearance of
long permutation cycles, which are necessary to see BEC. At $T =
0.2$ K, the large zero-point motion of the $^4$He atoms makes the
vacancy delocalized and undetectable inside the crystal,
which looks like a commensurate system. Since the number of
lattice sites is different from the number of particles, this
means that different polymers may superpose over the same lattice
site: this occurrence strongly enhances the possibility for the
atoms to permutate and allows the creation of long permutation
cycles which close on periodic boundary conditions. The appearance
of configurations presenting a non zero winding number, as the one
shown in Fig. \ref{SupersolidPicture}, indicates that the simulated
crystals below $T_0 = 0.2$ K support superfluidity. However, it is
not possible to give a reliable estimation for the superfluid
density $\rho_s/\rho$ in these systems, since the smallest value
for $\rho_s/\rho$ computable with the winding number estimator is
of the order of 1\%, that is of the same order of the value
expected from the experimental measurements.

In order to study how the vacancy concentration in quantum
solids affect the onset temperature of BEC, we have computed the
one-body density matrix also for fcc $^4$He crystals with $X_v =
1/108$, $X_v = 1/128$, and $X_v = 1/256$. It is worth noticing that the result 
for $X_v= 1/128$ has been obtained in
a simulation with two vacancies in a lattice of 256 site points and the
results for both the condensate fraction and onset temperature follow the
same $X_v$ dependence that the single vacancy cases.
In Table \ref{Tab_T0n0}, we show the
onset temperature of BEC $T_0$ and the condensate fraction $n_0$
at low temperature obtained with PIMC in the three crystals we
have studied. We notice that, for the lowest $X_v$, we
get $T_0 = 0.15 \pm 0.05$, which is close to the
temperatures at which supersolidity has been experimentally
observed.

\begin{table}
 \newcolumntype{C}{>{\centering\arraybackslash}X}
 \begin{tabularx}{0.8\textwidth}{cccCC}
 \hline
 \hline
  $N_v$ \hspace{20pt}& $N_s$ \hspace{20pt}& $X_v$ \hspace{20pt}& $T_0$ (K) & $n_0$ \\
  \hline
  $1$ \hspace{20pt}& $108$ \hspace{20pt}& $1/108$ \hspace{20pt}& $0.50 \pm 0.10$ & $(1.81 \pm 0.14) 10^{-3}$ \\
  $2$ \hspace{20pt}& $256$ \hspace{20pt}& $1/128$ \hspace{20pt}& $0.40 \pm 0.075$ & $(1.09 \pm 0.13) 10^{-3}$ \\
  $1$ \hspace{20pt}& $180$ \hspace{20pt}& $1/180$ \hspace{20pt}& $0.20 \pm 0.05$ & $(9.0 \pm 0.8) 10^{-4}$ \\
  $1$ \hspace{20pt}& $256$ \hspace{20pt}& $1/256$ \hspace{20pt}& $0.15 \pm 0.05$ & $(7.2 \pm 0.8) 10^{-4}$ \\
  \hline
  \hline
\end{tabularx}
\caption{The onset temperature of BEC $T_0$ and the condensate fraction $n_0$ at 
low temperature as a function of the vacancy concentration $X_v = N_v/N_s$, $N_v$ and $N_s$ being the number of vacancies and of lattice sites, respectively.}\label{Tab_T0n0}
\end{table}

\begin{figure}
\includegraphics[width=0.7\linewidth]{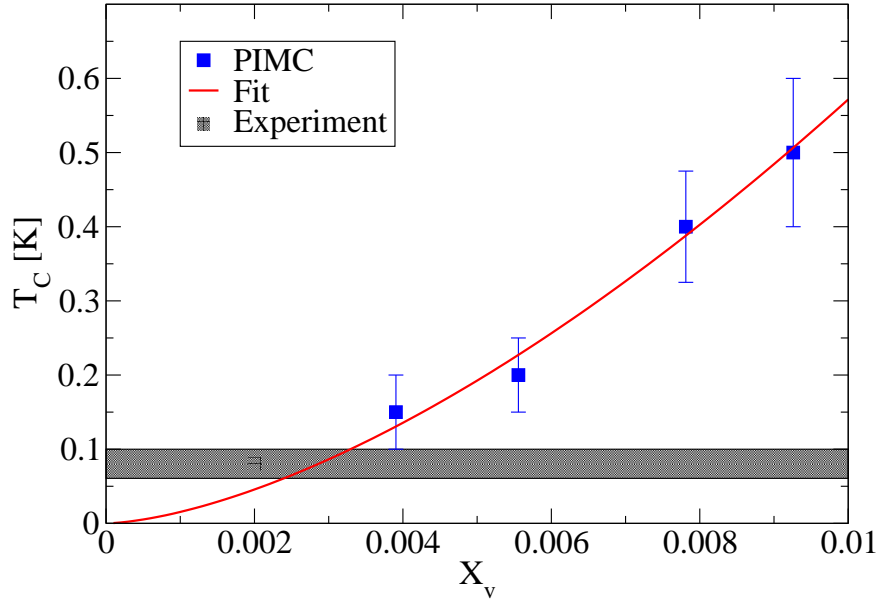}
\caption{(Color online) The onset temperature $T_0$ of BEC in
incommensurate solid $^4$He as a function of the vacancy
concentration $X_v$: the blue squares represent the results
obtained with the PIMC method; the red line is a fit to the data with a 
power law $T_0 = A X_v^B$ with optimal values $A=789$ K and $B=1.57$; the grey band
indicates the temperature at which the NCRI appears experimentally.}\label{T0_Xv}
\end{figure}

In Fig. \ref{T0_Xv}, we plot our results for $T_0$ as a function of $X_v$.
Our results for $T_0$ do not follow the law $T_0
\sim X_v^{2/3}$, obtained from a description of solid $^4$He in terms
of a  rarefied Gross-Pitaevskii superfluid gas of vacancies, 
as proposed by Anderson in  Ref. \cite{Anderson09}. This seems to suggest that, at least in
the range of $X_v$ we have  been able to study, the correlations between vacancies have an important effect on $T_0$ and the system cannot be
described within a mean-field approach. Nonetheless, our qualitative
description of $^4$He crystals supports the hypothesis~\cite{Anderson09} according to which it is not reasonable to regard vacancies in
quantum solids as strictly local entities.

In an attempt to estimate which should be the
vacancy concentration in
$^4$He crystals needed to have BEC appearing at the temperature $T_c$
measured experimentally for the supersolid transition, we have plotted in
Fig. \ref{T0_Xv} a power function trying to fit the PIMC results. 
According to this empirical law, $^4$He crystals  with a vacancy concentration $X_v \sim
2-3 \times 10^{-3}$ would have an onset temperature $T_0$ in agreement with the
experimental values $T_c$. This result for $X_v$ is in good agreement with the
equilibrium vacancy concentration in solid $^4$He at zero temperature obtained variationally with the shadow
wave function~\cite{Pessoa09}.

In conclusion, we have shown that the onset temperature $T_0$ of BEC in $^4$He
crystals presenting vacancies, calculated using
the PIMC method, is comparable with
the experimental measurements of the supersolid transition temperature when
the concentration of vacancies is small enough ($X_v \sim
2-3 \times 10^{-3}$). PIMC simulations also show clearly that when this
onset temperature is reached the vacancies become completely delocalized objects, as hypothesized in the 
past~\cite{Burns,Anderson09} 
and never microscopically observed so far.

This work was partially supported by DGI (Spain) under Grant No.
FIS2008-04403 and Generalitat de Catalunya under Grant No. 2009-SGR1003.

\end{document}